\documentstyle[12pt]{article}
\def\be {\begin{equation}}
\def\ee {\end{equation}}
\def\ba {\begin{eqnarray}}
\def\ea {\end{eqnarray}}

\def\bi {\begin{itemize}}
\def\ei {\end{itemize}}
\begin{document}
\def\bea{\begin{eqnarray}}
\def\eea{\end{eqnarray}}
\title{\bf {Holographic modified gravity}}
 \author{M.R. Setare  \footnote{E-mail: rezakord@ipm.ir}
  \\ {Department of Science,  Payame Noor University. Bijar, Iran}}
\date{\small{}}

\maketitle
\begin{abstract}
In this paper we study cosmological application of holographic dark
energy density in the modified gravity framework. We employ
 the holographic model of dark energy to obtain the equation of state  for the holographic energy density
 in spatially flat universe. Our calculation show, taking $\Omega_{\Lambda}=0.73$ for
the present time, it is possible to have $w_{\rm \Lambda}$ crossing
$-1$. This implies that one can generate phantom-like equation of
state from a holographic dark energy model in flat universe in the
modified gravity cosmology framework. Also we develop a
reconstruction scheme for the modified gravity with $f(R)$ action.
 \end{abstract}

\newpage

\section{Introduction}
Nowadays it is strongly believed that the universe is experiencing
an accelerated expansion. Recent observations from type Ia
supernovae \cite{SN} in associated with Large Scale Structure
\cite{LSS} and Cosmic Microwave Background anisotropies \cite{CMB}
have provided main evidence for this cosmic acceleration. In order
to explain why the cosmic acceleration happens, many theories have
been proposed. It is the most accepted idea that a mysterious
dominant component, dark energy, with negative pressure, leads to
this cosmic acceleration, though its nature and cosmological origin
still remain enigmatic at present. An alternative proposal for dark
energy is the dynamical dark energy scenario. The cosmological
constant puzzles may be better interpreted by assuming that the
vacuum energy is canceled to exactly zero by some unknown mechanism
and introducing a dark energy component with a dynamically variable
equation of state. The dynamical dark energy proposal is often
realized by some scalar field mechanism which suggests that the
energy form with negative pressure is provided by a scalar field
evolving down a proper potential.\\
In recent years, many string theorists have devoted to understand
and shed light on the cosmological constant or dark energy within
the string framework. The famous Kachru-Kallosh-Linde-Trivedi (KKLT)
model \cite{kklt} is a typical example, which tries to construct
metastable de Sitter vacua in the light of type IIB string theory.
Furthermore, string landscape idea \cite{landscape} has been
proposed for shedding light on the cosmological constant problem
based upon the anthropic principle and multiverse speculation.
Although we are lacking a quantum gravity theory today, we still can
make some attempts to probe the nature of dark energy according to
some principles of quantum gravity. The holographic dark energy
model is just an appropriate example, which is constructed in the
light of the holographic principle of quantum gravity theory. That
is to say, the holographic dark energy model possesses some
significant features of an underlying theory of dark energy.
Currently, an interesting attempt for probing the nature of dark
energy within the framework of quantum gravity is the so-called
``holographic dark energy'' proposal
\cite{Cohen:1998zx,Horava:2000tb,Hsu:2004ri,Li:2004rb}. It is well
known that the holographic principle is an important result of the
recent researches for exploring the quantum gravity (or string
theory) \cite{holoprin}. This principle is enlightened by
investigations of the quantum property of black holes. Roughly
speaking, in a quantum gravity system, the conventional local
quantum field theory will break down. The reason is rather simple:
For a quantum gravity system, the conventional local quantum field
theory contains too many degrees of freedom, and such many degrees
of freedom will lead to the formation of black hole so as to break
the effectiveness of the quantum field theory.

For an effective field theory in a box of size $L$, with UV cut-off
$\Lambda$ the entropy $S$ scales extensively, $S\sim L^3\Lambda^3$.
However, the peculiar thermodynamics of black hole \cite{bh} has led
Bekenstein to postulate that the maximum entropy in a box of volume
$L^3$ behaves nonextensively, growing only as the area of the box,
i.e. there is a so-called Bekenstein entropy bound, $S\leq
S_{BH}\equiv\pi M_P^2L^2$. This nonextensive scaling suggests that
quantum field theory breaks down in large volume. To reconcile this
breakdown with the success of local quantum field theory in
describing observed particle phenomenology, Cohen et al.
\cite{Cohen:1998zx} proposed a more restrictive bound -- the energy
bound. They pointed out that in quantum field theory a short
distance (UV) cut-off is related to a long distance (IR) cut-off due
to the limit set by forming a black hole. In other words, if the
quantum zero-point energy density $\rho_{\Lambda}$ is relevant to a
UV cut-off $\Lambda$, the total energy of the whole system with size
$L$ should not exceed the mass of a black hole of the same size,
thus we have $L^3\rho_{\Lambda}\leq LM_P^2$. This means that the
maximum entropy is in order of $S_{BH}^{3/4}$. When we take the
whole universe into account, the vacuum energy related to this
holographic principle \cite{holoprin} is viewed as dark energy,
usually dubbed holographic dark energy. Such a holographic dark
energy looks reasonable, since it may provide simultaneously natural
solutions to both dark energy problems as demonstrated in
Ref.\cite{Li:2004rb}. The holographic dark energy model has been
tested and constrained by various astronomical observations
\cite{obs3}. Furthermore, the holographic dark energy model has been
extended to include the spatial curvature contribution, i.e. the
holographic dark energy model in non-flat space \cite{nonflat}.
 Because the holographic energy density belongs to a
dynamical cosmological constant, we need a dynamical frame to
accommodate it instead of general relativity. Therefore it is
worthwhile to investigate the holographic energy density in the
framework of the Brans-Dicke theory \cite{{gong},{mu},
{tor},{set1}}. Einstein's theory of gravity may not describe gravity
at very high energy. The simplest alternative to general relativity
is Brans-Dicke scalar-tensor theory \cite{bd}. Modified gravity
provides the natural gravitational alternative for dark energy
\cite{odi1}. Moreover, modified gravity present natural unification
of the early-time inflation and late-time acceleration thanks to
different role of gravitational terms relevant at small and at large
curvature. Also modified gravity may naturally describe the
transition from non-phantom phase to phantom one without necessity
to introduce the exotic matter. But among the most popular modified
gravities which may successfully describe the cosmic speed-up is
$F(R)$ gravity. Very simple versions of such theory like $1/R$
\cite{1} and $1/R + R^2$ \cite{2} may lead to the effective
quintessence/phantom late-time universe (to see solar system
constraints on modified dark energy models refer to \cite{odin}).
Another theory proposed as gravitational dark energy is
scalar-Gauss-Bonnet gravity \cite{3} which is closely related with
low-energy string effective action.
\\
In present paper, using the holographic model of dark energy in
spatially flat universe, we obtain equation of state for holographic
dark energy density in framework  of modified gravity for a universe
enveloped by $R_h$ as the system's IR cut-off. The current available
observational data imply that the holographic vacuum energy behaves
as phantom-type dark energy, i.e. the equation-of-state of dark
energy crosses the cosmological-constant boundary $w=-1$ during the
evolution history. We show this phantomic description of the
holographic dark energy in flat universe with $0.21\leq c\leq 2.1$ .
Also we develop a reconstruction scheme for the modified gravity
with $f(R)$ action, the known holographic energy density is used for
this reconstruction.
\section{Modified gravity and holographic dark energy}
 The action of modified gravity is given
 by
\begin{equation}
S=\int d^{4}x[f(R)+L_{m}] . \label{action*}
\end{equation}
where $L_{m}$ is the matter Lagrangian density. The equivalent form
of above action is \cite{odi1}
\begin{equation}
S=\int d^{4}x \sqrt{-g}[P(\phi)R+Q(\phi)+L_{m}]. \label{action*1}
\end{equation}
where $P$ and $Q$ are proper functions of the scalar field $\phi$.
By the variation of the action (\ref{action*1}) with respect to the
$\phi$, we obtain
\begin{equation}
P'(\phi)R+Q'(\phi)=0 \label{action*2}
\end{equation}
which may be solved with respect to $\phi$:
\begin{equation}
\phi=\phi(R) \label{action*3}
\end{equation}
By the variation of the action (\ref{action*1}) with respect to the
metric $g_{\mu\nu}$, one can obtain
\begin{equation}
\frac{-1}{2}g_{\mu\nu}[P(\phi)R+Q(\phi)]-R_{\mu\nu}P(\phi)+\nabla_{\mu}\nabla_{\nu}P(\phi)-
g_{\mu\nu}\nabla^{2}P(\phi)+\frac{1}{2}T_{\mu\nu}=0
 \label{action*4}
\end{equation}
where $T_{\mu\nu}$ is the energy-momentum tensor. The equations
corresponding to standard spatially-flat FRW universe are
\begin{equation}
\rho=6H^2P(\phi)+Q(\phi)+6H\frac{dP(\phi)}{dt} \label{action*5}
\end{equation}
\begin{equation}
p=-(4\dot{H}+6H^2)P(\phi)-Q(\phi)-2\frac{d^{2}P(\phi)}{dt^2}-4H\frac{dP(\phi)}{dt}
 \label{action*6}
\end{equation}
where, $p$ and $\rho$ are the pressure and energy density due to the
scalar field in the modified gravity framework. By combining
(\ref{action*5}) and (\ref{action*6}) and deleting $Q(\phi)$, we
find
\begin{equation}
p+\rho=-2\frac{d^{2}P(\phi)}{dt^2}+2H\frac{dP(\phi)}{dt}-4\dot{H}P(\phi)
 \label{action*7}
\end{equation}
 Now we suggest a
correspondence between the holographic dark energy scenario and the
above modified dark energy model. The holographic energy density
$\rho_{\Lambda}$ is chosen to be \be
\rho_{\Lambda}=\frac{3c^2}{R_{h}^2} \label{holo}\ee where $c$ is a
constant, and $R_h$ is the future event horizon given by \be
\label{rh}
  R_h= a\int_t^\infty \frac{dt}{a}=a\int_a^\infty\frac{da}{Ha^2}
 \ee
 The critical energy density, $\rho_{cr}$, is given by following relation
\begin{eqnarray} \label{ro}
\rho_{cr}=3H^2
\end{eqnarray}
Now we define the dimensionless dark energy as \be
\Omega_{\Lambda}=\frac{\rho_{\Lambda}}{\rho_{cr}}=\frac{c^2}{R_{h}^2H^2}\label{omega}
\ee Using definition $\Omega_\Lambda$ and relation (\ref{ro}),
$\dot{R_{h}}$ gets: \be \label{ldot} \dot{R_{h}} =
R_{h}H-1=\frac{c}{\sqrt{\Omega_\Lambda}}-1,
\end{equation}
Let us consider the dark energy dominated universe. In this case the
dark energy evolves according to  its conservation law \be
\dot{\rho}_{\Lambda}+3H(\rho_{\Lambda}+P_{\Lambda})=0
\label{coneq}\ee By considering  the definition of holographic
energy density $\rho_{\rm \Lambda}$, and using Eq.(\ref{ldot})one
can find:
\begin{equation}\label{roeq}
\dot{\rho_{\Lambda}}=\frac{-2}{R_{h}}(\frac{c}{\sqrt{\Omega_\Lambda}}-1)\rho_{\Lambda}
\end{equation}
Substitute this relation into Eq.(\ref{coneq})  we obtain
\begin{equation}\label{stateq}
w_{\rm \Lambda}=-(\frac{1}{3}+\frac{2\sqrt{\Omega_{\rm
\Lambda}}}{3c}).
\end{equation}
A direct fit of the present available SNe Ia data with this
holographic model indicates that the best fit result is $c=0.21$
\cite{HG}. Recently, by calculating the average equation of state of
the dark energy and the angular scale of the acoustic oscillation
from the BOOMERANG and WMAP data on the CMB to constrain the
holographic dark energy model, the authors show that the reasonable
result is $c\sim 0.7$ \cite{cmb1}. In the other hand, in the study
of the constraints on the dark energy from the holographic
connection to the small $l$ CMB suppression, an opposite result is
derived, i.e. it implies the best fit result is $c=2.1$ \cite{cmb3}.
Thus according to these studies $0.21\leq c\leq 2.1$.  Taking
$\Omega_{\Lambda}=0.73$ for the present time, in the case of
$c=0.21$, we obtain $w_{\rm \Lambda}=-3.04$, in the other hand for
$c=2.1$, one can obtain, $w_{\rm \Lambda}=-0.6$. Using
Eq.(\ref{stateq}), one can see that by considering $c\leq
\sqrt{\Omega_{\Lambda}}$ we obtain $w_{\rm \Lambda}\leq -1$.
Therefore taking $\Omega_{\Lambda}=0.73$ for the present time, it is
possible to have $w_{\rm \Lambda}$ crossing $-1$.
\\
As one can redefine the scalar field $\phi$ properly, we may choose
\begin{equation}\label{stateq1}
\phi=t.
\end{equation}
Now using Eqs.(\ref{holo}), (\ref{stateq}), one can rewrite
Eq.(\ref{action*7}) as
\begin{equation}
2\frac{d^{2}P(t)}{dt^2}-2H\frac{dP(t)}{dt}+4\dot{H}P(t)+2\Omega_{\Lambda}H^2(1-\frac{\sqrt{\Omega_{\Lambda}}}{c})=0
 \label{action*8}
\end{equation}
In principle, by solving Eq.(\ref{action*8}) we find the form of
$P(\phi)$. Using Eqs. (\ref{action*5}), (\ref{holo}), we also find
the form of $Q(\phi)$ as
\begin{equation}
Q(\phi)=3\Omega_{\Lambda}H^2-6H^2 P(\phi)-6H\frac{dP(\phi)}{dt}
 \label{action*9}
\end{equation}
\section{Modified gravity and its
 reconstruction from the holographic dark energy}
In this section we consider another approach \cite{odi2} to
realistic cosmology in holographic modified gravity. We start with
general $f(R)$-gravity action (\ref{action*}) but without the matter
term. For the spatially flat FRW universe we have
\begin{equation}
\rho=f(R)-6(\dot{H}+H^2-H\frac{d}{dt})f'(R)
 \label{action*10}
\end{equation}
\begin{equation}
p=f(R)-2(-\dot{H}-3H^2+ \frac{d^2}{dt^2}+2H\frac{d}{dt})f'(R)
 \label{action*11}
\end{equation}
where
\begin{equation}
R=6\dot{H}+12H^2
 \label{action*12}
\end{equation}
Again we use the holographic dark energy density and substitute
Eq.(\ref{holo}) into Eq.(\ref{action*10})
\begin{equation}
3\Omega_{\Lambda}H^2=f(R)-6(\dot{H}+H^2-H\frac{d}{dt})f'(R)
 \label{action*13}
\end{equation}
thus
\begin{equation}
f(R)=3\Omega_{\Lambda}H^2+6(\dot{H}+H^2-H\frac{d}{dt})f'(R)
 \label{action*14}
\end{equation}
Using Eqs.(\ref{holo}), (\ref{stateq}), and substituting $f(R)$ into
Eq.(\ref{action*11}) one can obtain
\begin{equation}
2\frac{d^2}{dt^2}f'(R)-2H\frac{d}{dt}f'(R)+4\dot{H}f'(R)+2\Omega_{\Lambda}H^2(1-\frac{\sqrt{\Omega_{\Lambda}}}{c})=0
 \label{action*15}
\end{equation}
or in another form
\begin{equation}
2(f'''\dot{R}^{2}+f''\ddot{R})-2Hf''\dot{R}+4\dot{H}f'+2\Omega_{\Lambda}H^2(1-\frac{\sqrt{\Omega_{\Lambda}}}{c})=0
 \label{action*16}
\end{equation}
We shall consider the following simple solution
\begin{equation}\label{112}
a=a_0(t_s-t)^{h_0},
\end{equation}
where   $a_0$, $h_0$ and  $t_{s}$ are constant. Substituting
Eq.(\ref{112}) into Eq.(\ref{action*12}), give us following relation
for scalar curvature
\begin{equation}
R=\frac{12h_{0}^{2}-6h_0}{(t_s-t)^{2}}
 \label{action*122}
\end{equation}
Using Eqs.(\ref{rh}, \ref{112}) we can write
\begin{equation}\label{116}
R_{h}=a_0(t_s-t)^{h_0}\int_{t}^{t_s}\frac{dt}{a_0(t_s-t)^{h_0}}=\frac{t_s-t}{1-h_0}
\end{equation}
Now using definition $\rho_{\Lambda}$ and above relation we obtain
the time behaviour of holographic dark energy as
\begin{equation}\label{117}
\rho_{\Lambda}=\frac{3c^2}{R_{h}^2}=\frac{3c^2(1-h_0)^{2}}{(t_s-t)^{2}}
\end{equation}
Substituting the above $\rho_{\Lambda}$ into Eq.(\ref{action*10}),
and using Eqs.(\ref{112},\ref{action*122}) one can obtain
\begin{equation}\label{118}
\frac{72h_{0}^{2}(1-2h_0)}{(t_s-t)^{4}}f''(R)-\frac{6h_0(h_0-1)}{(t_s-t)^{2}}f'(R)+f(R)=
\frac{3c^2(1-h_0)^{2}}{(t_s-t)^{2}}
\end{equation}
Again we use Eq.(\ref{action*122}) and rewrite the above
differential equation as following
\begin{equation}\label{119}
f''(R)+\frac{a}{R}f'(R)+\frac{b}{R^2}f(R)= \frac{d}{R}
\end{equation}
where
\begin{equation}\label{120}
a=\frac{h_0-1}{2}, \hspace{1cm} b=\frac{1-2h_0}{2}, \hspace{1cm}
d=\frac{c^2(1-h_0)^{2}}{4h_0}
\end{equation}
The solution of differential equation (\ref{119}) is given by
\begin{equation}\label{121}
f(R)=C_{1}R^{\frac{1}{2}\left(\frac{3-h_0}{2}-\sqrt{\frac{(h_0-3)^{2}}{4}+4h_0-2}\right)}+
C_{2}R^{\frac{1}{2}\left(\frac{3-h_0}{2}+\sqrt{\frac{(h_0-3)^{2}}{4}+4h_0-2}\right)}+\frac{c^2(1-h_0)^{2}R}{2h_{0}^{2}}
\end{equation}
where $C_{1},C_{2}$ are constant. Therefore, a consistent modified
gravity whit holographic dark energy in flat space has the above
form. Let us recall the two sufficient conditions which often lead
to realistic models \cite{{odin},{hu}}
\begin{equation}\label{122}
\lim_{R\rightarrow 0} f(R)=0,
\end{equation}
this condition ensures the disappearance of the cosmological
constant in the limit of flat space-time. One can see that
(\ref{121}) simply satisfy the above condition. In order that the
accelerating expansion in the present universe could be generated,
let us consider that $f(R)$ could be a small constant at present
universe, that is,
\begin{equation}\label{123}
f(R_0)=-2R_0, \hspace{1cm} f'(R_0)\sim 0,
\end{equation}
where $R_0\sim (10^{-33}eV)^{2}$ is current curvature \cite{odin}.
By impose the conditions (\ref{123}) on the solution (\ref{121}) we
can obtain the constants $C_{1}$ and $C_{2}$ as following
\begin{equation}\label{124}
C_1=\frac{-R_{0}^{1-u}}{u+v}[2v+(v+1)\frac{c^{2}(1-h_{0})^{2}}{2h_{0}^{2}}]
\end{equation}
\begin{equation}\label{125}
C_2=\frac{R_{0}^{1-v}}{u+v}[2u-(1-u)\frac{c^{2}(1-h_{0})^{2}}{2h_{0}^{2}}]
\end{equation}
where
\begin{equation}\label{126}
u=\frac{1}{2}\left(\frac{3-h_0}{2}-\sqrt{\frac{(h_0-3)^{2}}{4}+4h_0-2}\right),
\hspace{0.3 cm}
v=\frac{1}{2}\left(\frac{3-h_0}{2}+\sqrt{\frac{(h_0-3)^{2}}{4}+4h_0-2}\right)
\end{equation}
The model Eq.(\ref{121}) is similar to the model of Eq.(12) in
\cite{odin2}, similarly our model also leads to acceptable cosmic
speed-up and is consistent with solar system tests.
\section{Conclusions}
Within the different candidates to play the role of the dark energy,
the modified gravity, has emerged as a possible unification of dark
matter and dark energy. In the present paper we have studied
cosmological application of holographic dark energy density in the
modified gravity framework. By considering the holographic energy
density as a dynamical cosmological constant, we have obtained the
equation of state for the holographic energy density in the modified
gravity framework. We have shown if $c\leq\sqrt{\Omega_{\Lambda}}$,
the holographic dark energy model also will behave like a phantom
model of dark energy the amazing feature of which is that the
equation of state of dark energy component $w_{\rm \Lambda}$ crosses
$-1$. Hence, we see, the determining of the value of $c$ is a key
point to the feature of the holographic dark energy and the ultimate
fate of the universe as well. Finally we have developed a
reconstruction scheme for modified gravity with $f(R)$ action. We
have considered the energy density in Eq.(\ref{action*10}) in
holographic form, then by assumption a simple solution as
Eq.(\ref{112}) we could obtain a differential equation for $f(R)$,
the solution of this differential equation give us a modified
gravity action which is consistent with holographic dark energy
scenario.
\section{Acknowledgment}
The author would like to thank the referee because of his useful
comments, which assisted to prepare better frame for this study.

\end{document}